\documentclass[iop]{emulateapj}
\usepackage{natbib}
\usepackage{booktabs}
\usepackage{color}

\def\apjl{ApJL}               
\def\apjs{ApJS}               
\def\aap{A\&A}                

\newcommand{\stdFig}[4]{
\begin{figure}
\center
\includegraphics[scale=#4]{#1}
\caption{\small  #2 }
\label{#3}
\end{figure}
}

\newcommand{\stdFullFig}[4]{
\begin{figure*}
\center
\includegraphics[scale=#4]{#1}
\caption{\small  #2 }
\label{#3}
\end{figure*}
}

\newcommand{\stdTable}[4]{
\begin{table}
\label{#3}
\begin{center}
\bgroup 
\caption{#1}
\begin{small}
\renewcommand{\tabcolsep}{12pt}
\vspace{5 pt}
#2
\end{small}
\egroup
\end{center}
\end{table}
}

\newcommand{\stdFullTable}[4]{
\begin{table*}
\begin{center}
\bgroup 
\caption{#1}
\begin{small}
\renewcommand{\tabcolsep}{12pt}
\vspace{5 pt}
#2
\end{small}
\egroup
\end{center}
\label{#3}
\end{table*}
}

\newcommand{\msun}{\;{\rm M}_{\odot}}
\newcommand{\crs}{{\rm CR}}

\newcommand{\SB}{SB14} 
\newcommand{\HB}{HB12} 

\begin{document}

\title{Cosmological simulations of galaxy formation with cosmic rays}


\author{Munier Salem$^{1}$, Greg L. Bryan$^{1}$ and Cameron Hummels$^{2}$}
\altaffiltext{1}{Department of Astronomy, Columbia University, 550 West 120th Street, New York, NY 10027}
\altaffiltext{2}{University of Arizona. Department of Astronomy. Steward Observatory. 933 N Cherry Avenue, N309. Tucson, AZ}





\begin{abstract}
We investigate the dynamical impact of cosmic rays in cosmological simulations of galaxy formation using adaptive-mesh refinement simulations of a $10^{12} \msun$ halo. In agreement with previous work, a run with only our standard thermal energy feedback model results in a massive spheroid and unrealistically peaked rotation curves.  However, the addition of a simple two-fluid model for cosmic rays drastically changes the morphology of the forming disk.  We include an isotropic diffusive term and a source term tied to star formation due to (unresolved) supernova-driven shocks.  Over a wide range of diffusion coefficients, the CRs generate thin, extended disks with a significantly more realistic (although still not flat) rotation curve. We find that the diffusion of CRs is key to this process, as they escape dense star forming clumps and drive outflows within the more diffuse ISM.
\end{abstract}


\section{Introduction}

The formation of realistic disk galaxies in cosmological simulations has proven a considerable challenge over the years, largely due to the tendency for energetic feedback from star formation to be radiated away \citep[e.g.,][and references therein]{Navarro:1991p1002, Thacker:2000p1040, Abadi:2003p639, Governato:2007p1022, Hummels2012}.  Recently, however, significant progress has been made by a number of groups, producing thin disks with nearly flat rotation curves \citep[e.g.,][]{Governato2010, Guedes2011, Stinson2013b, Agertz2013, Aumer2013, Brooks2011, Marinacci2014}.  While methodology has varied, these successes have generally involves a subgrid model tuned to generate large winds, either explicitly, or via a technique to reduce radiative loses in dense, star-forming gas.  Although this approach is certainly useful, the detailed physical nature of the feedback in these models is imposed as a sub-grid model and generally tuned to match observations.  Complementary work to produce a sub-grid model from first principles is still in early stages -- preliminary efforts have found it remarkably difficult to drive highly mass-loaded winds from baryon-rich disks \citep[e.g.,][]{MacLow1999, Joung2009, Creasey2013}.

Here, we take a somewhat different approach, and explore the impact of cosmic-ray (CR) pressure on the formation of disks in cosmological simulations.  Although the detailed dynamics of cosmic-rays are complicated, we adopt a simple two-fluid approach that has been widely used, and which, we argue, captures the key effects of cosmic-ray dynamics.  We assume that CRs are accelerated by supernovae blast waves, retaining a significant fraction of the SN energy, that the CRs are tightly coupled to the thermal gas (via magnetic interactions) except for a diffusive term.  This work builds on previous efforts to model CR pressure in previous 1D and 3D galaxy models, and is a direct (cosmological) follow-on to our previous work \citep[][hereafter \SB]{paperI}.  Early work \citep{Breitschwerdt1991, Breitschwerdt1993, Zirakashvili1996, Everett2008} showed that CRs could drive significant winds in one-dimensional steady-state models, a conclusion that was extended to time-dependent cases by \citet{Dorfi2013}.  Full, three-dimensional galaxy simulations including CRs have only recently been explored \citep{Ensslin2007, Jubelgas2008, Siejkowski2014}, affirming that CRs can influence galactic structure but generally did not include streaming or diffusion, which turns out to be a key ingredient in driving winds \citep[][\SB]{Uhlig2012}.

In \SB, we simulated an idealized galactic disk, demonstrating that CRs with diffusion generically drove outflows, with mass-loading factors approaching unity.  The precise results depended mostly on the CR diffusion coefficient, but also had a dependency on the amount of energy injected. This result, first submitted in July 2013, was confirmed by two letters both submitted a month after \SB\ \citep{Booth2013, Hanasz2013}, the latter of which explicitly included magnetic fields and anisotropic diffusion.  In this paper, we apply our CR model to galaxy formation in a cosmological context, demonstrating that it has a dramatic effect on the disk dynamics, resulting in systems much closer to those observed than our standard (pure thermal) feedback model.

\section{Methodology}
\label{sec:methodology}

Our cosmological galaxy simulation is based very closely on \cite{Hummels2012}, hereafter~\HB, who performed a ``zoom'' simulation of a forming $1.2 \times 10^{12} M_\odot$ halo within a $(20 \; {\rm Mpc}/h)^3$ box with cosmological parameters from the WMAP5 year results \citep{Komatsu2009} (see Table \ref{tab:params}), evolved from $z = 99$ to the present day.  We zoom on the same relatively isolated $\sim 10^{12} M_\odot$ halo (denoted halo 26 in that work), using the adaptive mesh refinement code Enzo -- see \citet{Bryan2013} for a description, and the same initial conditions and refinement strategy.  In particular, we use a base grid of $128^3$ cells per side, two levels of initial refinement (resulting in a dark matter particle mass of $4.9 \times 10^6$ $\msun$) and a maximum of 9 levels of AMR, thus providing a maximum comoving resolution of 305 $h^{-1}$pc.


\stdFullTable{Parameters}{

\begin{tabular}{cccccccccc}
\toprule		
 \multicolumn{2}{c}{CR Physics}							& \multicolumn{2}{c}{SF / Feedback}				& \multicolumn{4}{c}{Cosmology}							& \multicolumn{2}{c}{Numerics}			\\
\cmidrule(r){1-2} 										\cmidrule(rl){3-4} 							\cmidrule(l){5-8} 										\cmidrule(l){9-10} 
$\kappa_\crs$		&\hspace{-1cm} 	$\{0,.3,1,3\} \times10^{28}$ cm$^2$/s	& $\epsilon_{\rm SF}$	& .01					& $\Omega_{0}$			& .258	& $h$ 		& .719	& $\Delta x_{\rm min}$	&	425 pc*	\\
$\gamma_\crs$		& 	$4/3$							& $\epsilon_{\rm SN}$ 	& $3\times10^{-6}$		& $\Omega_{\rm \Lambda}$	& .742	& $\sigma_8$	& .796	& size				&	20 Mpc*	\\
$c_{s, {\rm max}}$	&	$1000$ km/s						& $f_\crs$				& 0.3					& $\Omega_{\rm b}$			& 0.044	& $z_i$		& 99 		&					&			\\
\bottomrule
\label{tab:params}
\end{tabular}
}{Simulation parameters}

We integrate CR physics into Enzo via a two-fluid model \citep{Jun1994,Drury1985, Drury1986}, whose features, limitations and implementation are described in \SB. Briefly, this model assumes a relativistic population of a few GeV protons treated as an ideal gas with $\gamma_\crs = 4/3$. Inhomogeneities in the ISM's magnetic field  (not explicitly modeled) scatter the CR's motion, tying them to the thermal plasma except for a diffusion term modeled with a homogenous, scalar diffusion coefficient, $\kappa_\crs$. Bulk motions of the thermal gas transport the CRs and perform adiabatic work on the CR fluid. In turn, the CR fluid exerts a scalar pressure on the thermal ISM. This model neglects diffusion of the CRs in energy, non-adiabatic CR energy loss terms and any attempt to directly model magnetic fields. 

This prescription necessitated the inclusion in Enzo of a new conserved fluid quantity, $\epsilon_\crs$, evolved with the robust ZEUS hydro method~\citep{Stone1992}, as described and tested in \SB. That work explored various diffusion coefficients, $\kappa_\crs \in \left[ 0,10^{29} \right] $ cm$^2$/s. We again vary the scalar diffusion coefficient across runs, including $\kappa_\crs \in [ 0 , 3\times 10^{27}, 10^{28}, 3 \times 10^{28} ]$ cm$^2$/s. Within galactic disks, both CR propagation models and observations motivate a value of a few times $10^{28}$ cm$^2$/s \citep[e.g.,][]{Ptuskin2006, Ackermann2012,Strong1998,Tabatabaei2013}, athough a more detailed model would treat this coefficient as an anisotropic tensor dependent on details of the magnetic field and CR momentum distribution. 

The present work occurs over cosmological timescales, during which our comoving grid cells expand in size appreciably in physical units. This adiabatic expansion manifests as a decay in the fluid's energy density. For the ultra-relativistic ray gas, with $\gamma_\crs = 4/3$, the decay follows $\partial_t \epsilon_\crs = - (\dot{a}/a) \epsilon_\crs$ where $a(t)$ is the cosmic scale factor.

Our resolution fails to resolve the formation of molecular clouds and stars. Thus to capture this physics, we create collisionless ``star particles'' of mass $M_\star \ge 10^5 M_\odot$.  To determine the star formation rate, we follow the prescription of \cite{Cen1992}, updated as first described in \cite{OShea2004}, which essentially adopted a star formation rate $\dot{\rho_{\rm SF}} = \epsilon_{\rm SF} \rho / t_{\rm dyn}$.  Here $\epsilon_{\rm SF}$ is the star formation efficiency, and $t_{\rm din}$ is the dynamical time; a detailed explanation of parameters and their choices (listed in Table \ref{tab:params}) can be found in \HB, which employed an identical prescription without cosmic rays. 

We also include stellar feedback from Type II supernovae, which deposit a portion of the star particle's mass and energy back into the 
fluid quantities of the cell it occupies over a dynamical time, following
\begin{eqnarray}
\Delta M_{\rm gas}	&=& f_\star m_\star \\
\Delta E_{\rm gas}	&=&	(1 - f_\crs) \epsilon_{\rm SN} m_\star c^2 \\
\Delta E_{\rm CR }	&=&	f_\crs \epsilon_{\rm SN} m_\star c^2 \; ,
\end{eqnarray}
where $f_\star = 0.25$ is the mass fraction of the star ejected as winds and SN ejecta, $\epsilon_{\rm SN} = 3 \times 10^{-6}$ is the Type II supernovae efficiency and $f_\crs$ is the fraction of this energy feedback donated to the relativistic CR fluid. The choice of this latter CR feedback was explored in \SB. Following that work's fiducial run, we set $\epsilon_{\rm SN} = 3 \times 10^{-6}$.

\section{Results}
\label{sec:results}

\stdFig{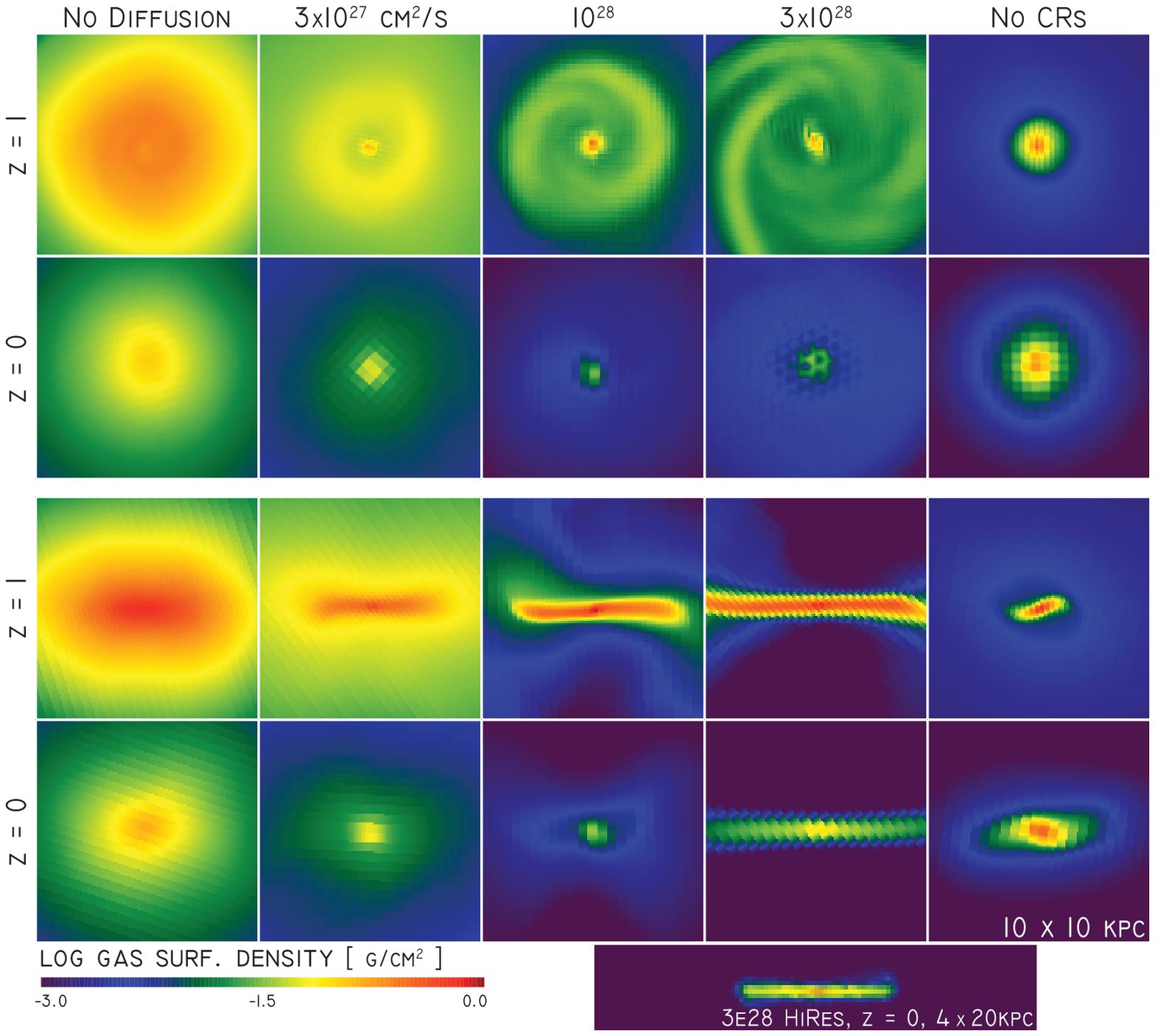}{Surface gas density at the center of our $\sim10^{12} M_\odot$ halo. Each column is a distinct simulation, where the cosmic ray fluid becomes increasingly diffusive from left to right. The rightmost column is a run devoid of CRs. The top two rows are a face-on view at redshifts 1 and 0, whereas the bottom two rows are edge-on. A $z=0$ view of a highly diffusive run with a factor of two higher spatial resolution (see text) at $z=0$ is shown in the bottom panel.}{fig:gasProj}{.5}

\stdFig{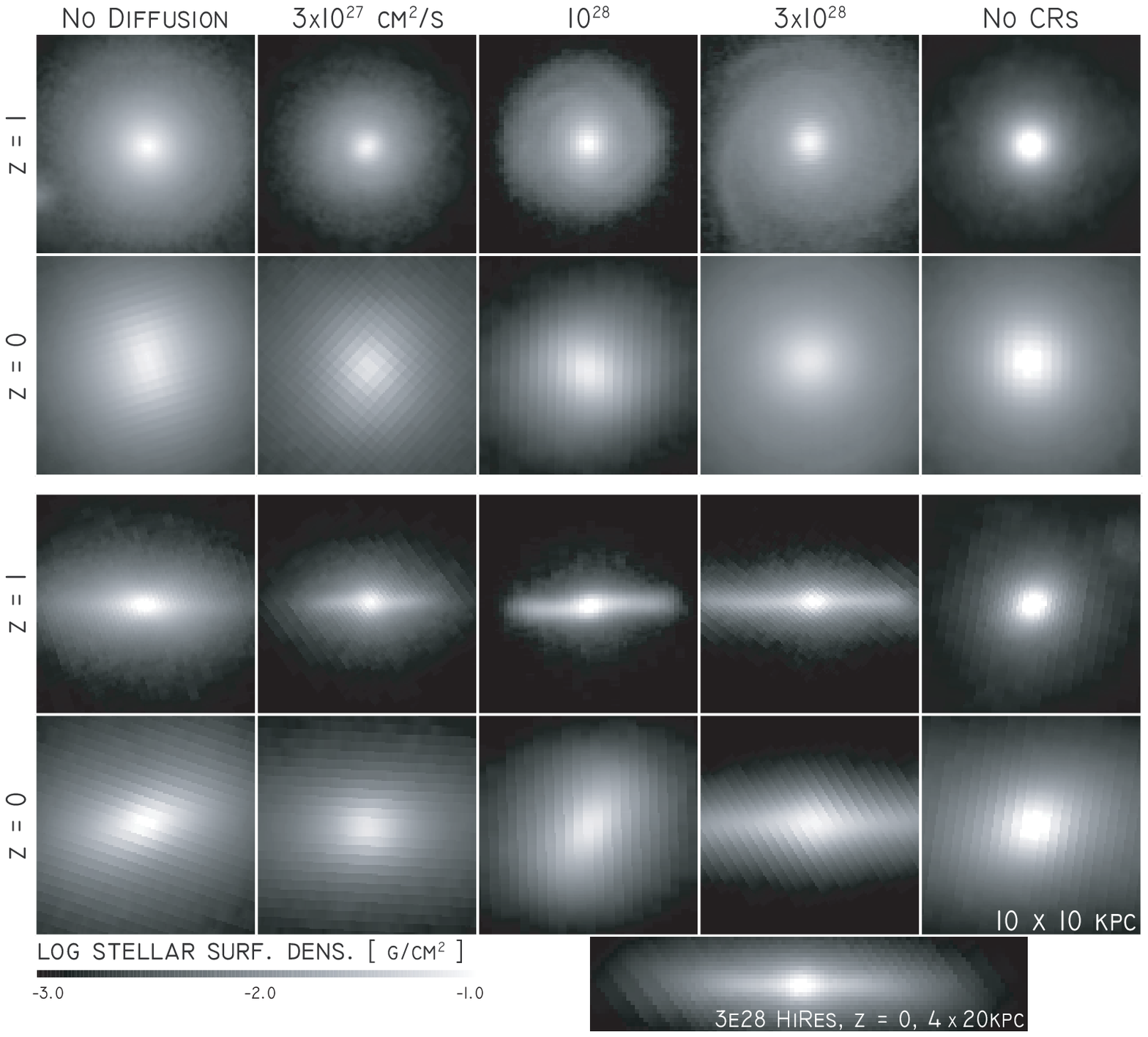}{Stellar surface density across our simulations in the center of our $\sim10^{12} M_\odot$ halo. See Figure \ref{fig:gasProj} for an explanation of layout. }{fig:starProj}{.5}

We begin by presenting a brief analysis of the distribution of baryons in our $1.2 \times 10^{12} M_\odot$ halo across our five simulations (four which vary the CR diffusion coefficient and a control run without CRs). Figures \ref{fig:gasProj} and \ref{fig:starProj} display gas and stellar column densities within a central 10 kpc$^2$ region at the center of the DM halo. The top two rows show face-on views of the baryonic disk at both $z=1$ and $z=0$, the former epoch featuring substantially higher physical resolution. The bottom two rows show the edge-on view. Each column of these figures represents a different simulation, where, from left to right, our cosmic rays run from zero-diffusion to very diffusive.

The rightmost column of Figures \ref{fig:gasProj} and \ref{fig:starProj} shows a run with the traditional (purely thermal) feedback of \cite{Cen1992} but no CR physics.  
Face- and edge-on views at $z=1$ show that this approach has concentrated the gas and stars within a massive, central, kpc-scale bulge. A rotationally supported disk of roughly 5 kpc in diameter is visible in the gaseous component, although this feature is mostly washed out in the stellar maps, particularly at $z = 0$.  As is well-known, purely thermal energy injection is not an efficient feedback mechanism for such simulations, as the energy is rapidly radiated away.

At the opposite extreme of behavior is our run with cosmic rays but no CR diffusion, plotted in the leftmost column of Figures \ref{fig:gasProj} and \ref{fig:starProj}. \SB~showed in an isolated galaxy that the pressure support of the CRs injected in SF regions led to a puffed up disk with the lowest SFRs among all runs considered but failed to launch any galactic winds to redistribute the baryons. In this cosmological setting we find these earlier results corroborated. Like the no-CR run, this no-diffusion simulation has failed to produce a thin disk of stars down to $z=0$, although it did manage to lower the resulting stellar mass by $\sim30\%$, lowering the rotation curve peak by over $100$ km/s. Finally the baryon fraction throughout the central 10 kpc has been lowered by $\sim10\%$, though this is somewhat due to a higher concentration of DM at the center, where less SF (and thus local feedback) has occurred compared to the no-CR run.

Our final three runs include not only CRs but also CR diffusion, shown in the middle three columns of Figures \ref{fig:gasProj} and \ref{fig:starProj}. These runs feature thin gaseous disks from high redshift down past $z=1$, complete with large, coherent spiral features in the more diffusive runs at $z=1$. By $z=0$, the disks are gas poor, with the ``middle''  $\kappa_\crs = 10^{28}$ cm$^2$/s run showing the least residual gas. 
The ratio of CR pressure to gas pressure in the disk varies mostly between a factor of roughly unity and ~100 for these runs.  At a radius of 10 kpc from the disk center, the CR energy density is about $4 \times 10^{12}$ erg cm$^{-2}$, within a factor of two of the value observed in the solar neighborhood.  This value is relatively constant across our simulations with non-zero diffusion coefficients.

In the stellar maps, we see all three of these runs have extended, thin stellar disks, with the thickness decreasing for more diffusive runs, to a fraction of a kiloparsec (although the over-sized bulge component likewise grows for the more diffusive runs, as local feedback from the CRs becomes less effective). These disks are somewhat thickened and ``washed out'' by $z=0$. Because our simulations take place in comoving coordinates, physical resolution degrades throughout the run. Thus for the $3\times10^{28}$ run, we performed an additional simulation where its maximum allowed refinement is enhanced by a factor of two after $z=1$ (the mass resolution is unchanged). This final, higher resolution run is shown edge-on at the bottom of Figures \ref{fig:gasProj} and \ref{fig:starProj}, and we see that the thin disk is recovered thanks to the higher resolution.

Figure \ref{fig:massRot} quantifies the mass distribution of the simulations at $z=0$, plotting radial profiles of enclosed gas, stellar and DM mass, as well as the cumulative baryon fraction and implied rotation curve.  For the no-CR run (black lines), this plot shows that the high concentration of baryons in the central kiloparsec has lead to a rotation curve peaked at roughly $700$ km/s. This run's behavior is consistent with the feedback-inclusive results of \HB.  

\SB~found that runs including CR diffusion, over a very wide range of coefficients, managed to launch winds from the disk with mass loading factors (normalized by the star formation rate) of order unity. Their less diffusive runs lifted more mass from the disk, ultimately processing more ISM into the CGM than into stars. Figure \ref{fig:slices} shows edge-on cutaway ``slices'' of gas density and CR energy density for our halos at $z=1$, with poloidal velocity vectors superimposed showing transport of the two-fluid.  All three diffusion-inclusive runs (middle three columns) feature robust flows above and below the disk, with velocities well above 100 km/s. 
For the most diffusive run, the speed of these winds exceeds 1000 km/s, although the flows become increasingly evacuated, leading to a lower mass flux. These flows first appear at substantially higher redshift and persist down to $z=0$ (indeed, they are stronger at higher redshift).

From Figure \ref{fig:massRot}, we find substantially lower stellar masses throughout the halo for our diffusion-inclusive runs (blue, green and red). The run most effective at suppressing star formation is the least diffusive $\kappa_\crs = 3 \times 10^{27}$ cm$^2$/s run, consistent with the results of \SB. This run produced roughly half as many stars as the no-CR run. All three diffusive runs perform better in this respect than the non-diffusive case. All CR-inclusive runs lower the cumulative baryon fraction by $30\%$ or more in the central 10 kpc, with the less diffusive runs decreasing the fraction at larger radii even more. Finally, these runs were also the most effective at driving down the rotation curve peak, with the least diffusive run dropping below $400$ km/s.

\stdFig{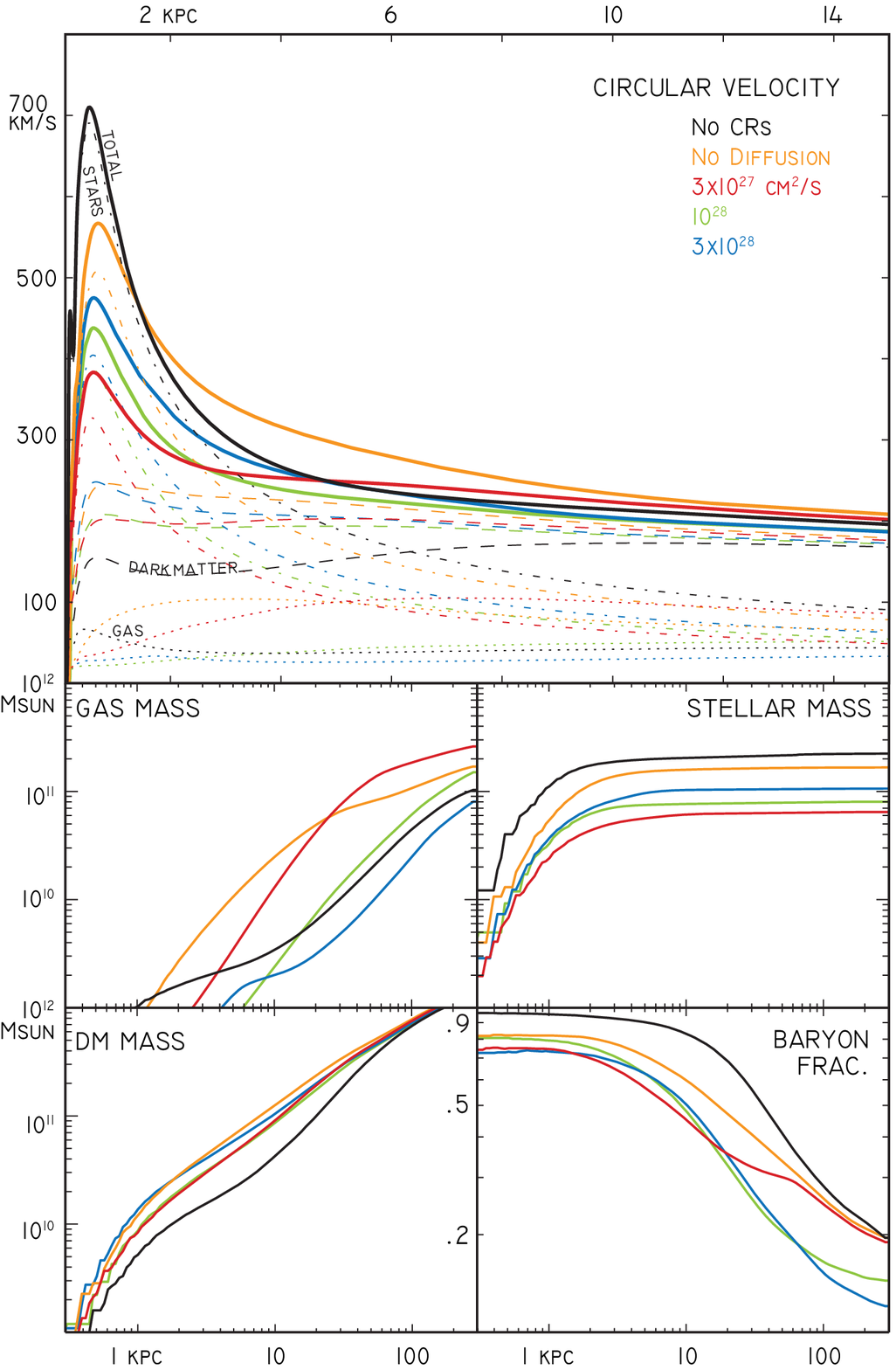}{{\it top}: 
Rotation curves for our five simulations at $z=0$. Thick solid lines denote $v = \sqrt{GM/r}$ including all mass components. The thin lines break down this rotation curve by mass, for the stars (dot-dash), gas (dotted) and dark matter (dashed).
{\it bottom:} 
Cumulative radial mass profiles across the same five runs, centered on our $\sim10^{12} M_\odot$ halo, for the baryonic gas, stars and dark matter, and the cumulative baryon fraction over for the same region. 
}{fig:massRot}{0.55}

\stdFullFig{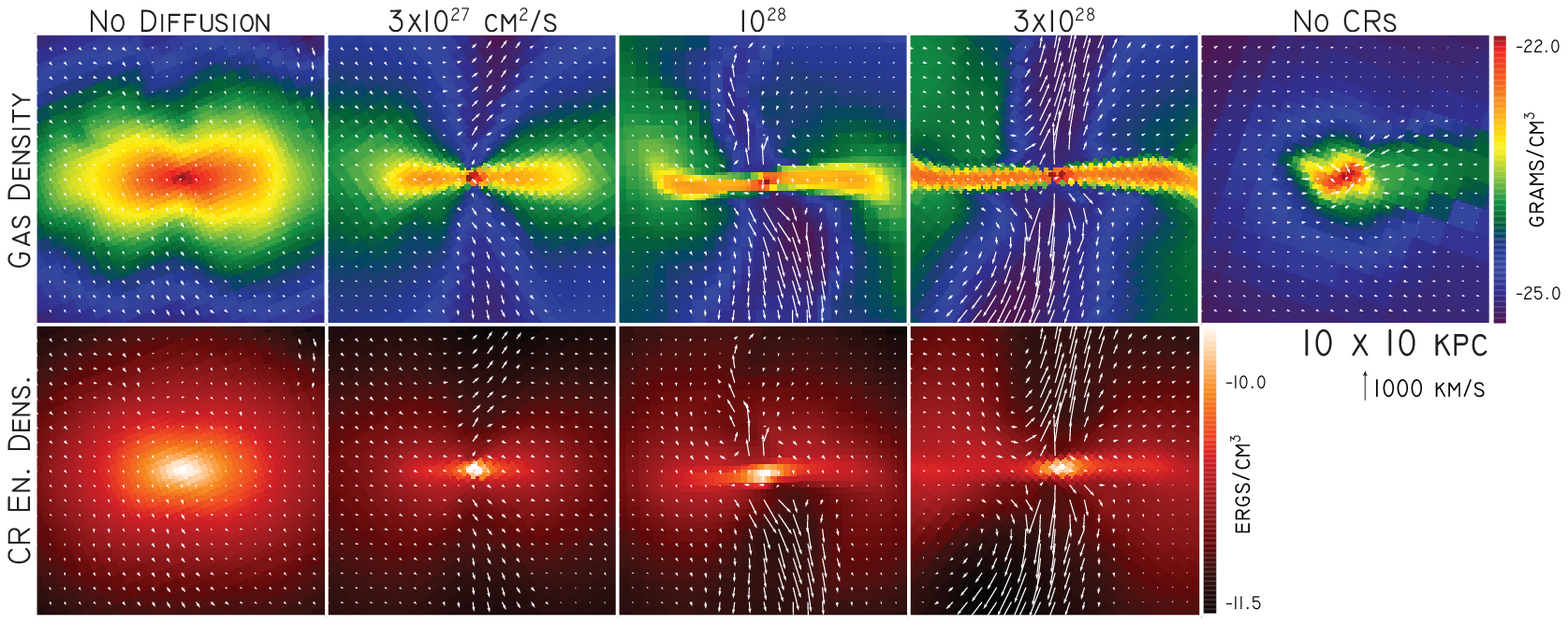}{{\it top:} edge-on cutaways of physical gas density with velocity vectors over-plotted across our five runs. Diffusive CR runs feature bipolar bulk flows of material away from the central disk. {\it bottom}: The same cutaways, but showing cosmic ray energy density.}{fig:slices}{1.0}

\section{Conclusion}
\label{sec:conclusion}

Our diffusive CR model successfully produces thin, extended baryonic disks that persist to low redshift, with fewer stars, a lower baryon fraction and a less pronounced peak in their rotation curves than in our thermal-only feedback models.  The details depend on the precise diffusion coefficient chosen, but, the overall result holds over a wide range of coefficients.  Although CR physics dramatically improves the realism of the resulting disks, these runs could not completely disband the unnaturally large spheroid of gas and stars at the center of the simulated galaxy, as the reflected in their peaked rotation curves. In fact, the runs with the most extended disks (the most diffusive $\kappa_\crs = 3 \times 10^{28}$ cm$^2$/s run) were also the runs with the largest central bulge. This suggests the precise role of CRs in feedback, and the spatial scales on which they are relevant, depend on the details of CR streaming and diffusion.  A more diffusive CR gas may play a less critical role in small scale feedback, on the level of SF clumps, than in global flows. This is in agreement with the higher resolution work of ~\SB, where strongly mass loaded winds arose from the disk, but failed to regulate fragmentation and star formation. This problem is likely due to the absence of other feedback processes, such as radiation and stellar winds, that play an important role in dense molecular gas which are unaffected by cosmic-ray feedback.  We also may be impacted by the relatively poor resolution of these cosmological runs, which still do not resolve the disk scale-height. 

In galaxy formation, ``feedback'', in its broadest sense, has come to connote disruption, expansion and redistribution on many scales, from processes that disrupt star-forming regions, to global flows of material within and beyond massive halos. The smallest scale feedback, necessary to arrest star formation on small-scales, clearly involves processes that are unresolved by today's generation of cosmological galaxy simulations.  This has led to ``sub-grid'' prescriptions that inject mass, momentum and energy back into the ISM, both in order to regular star formation {\it and} to generate large-scale winds, necessary to reduce the observed baryon fraction in galaxies.  An implicit assumption has often been made in these efforts: that these same physical processes operating on the sub-grid level, built together in aggregate, will drive global flows of material on large scales.  Our work suggests that sub-grid prescriptions that either directly launch winds or delay radiative losses may not be necessary to drive such outflows, and that the acceleration of these flows may occur on scales which can be resolved in such simulations, as long as the proper physical processes are included.  Our flows do not feature dense clumps rising ballistically through a more rarefied medium. Instead, as shown in the higher resolution runs of~\SB, we see a bulk transport of the multiphase ISM rising buoyantly, and \emph{gaining speed}, beyond the sites of star formation. The flows are primarily driven not by a hot, evacuated cavity of thermal gas, but by the pressure support of a relativistic component that is everywhere intermixed with the fluid. 

\section*{Acknowledgments}

We acknowledge financial support from NSF grants AST-0908390 and AST-1008134, and NASA grant NNX12AH41G,  as well as computational resources from NSF XSEDE, and Columbia University's Yeti cluster. 

\bibliography{ms}{}

\end{document}